\documentclass[10pt]{bmc_article}

\usepackage{cite} 
\usepackage{url}  
\usepackage{ifthen}  
\usepackage{multicol}   
\usepackage[utf8]{inputenc} 
\usepackage{epsfig}
\usepackage{here}
\usepackage{subfigure}
\usepackage{multirow}
\usepackage{amsfonts}
\newcommand{\urlBiBTeX}[1]{\url{#1}}
\renewcommand*{\thefootnote}{\fnsymbol{footnote}}

\newboolean{publ}

\newenvironment{bmcformat}{\baselineskip20pt\sloppy\setboolean{publ}{false}}{\baselineskip20pt\sloppy}

\begin{document}
\begin{bmcformat}


\title{Preventing common hereditary disorders through time-separated twinning}


\author{Alexander Churbanov\correspondingauthor$^1$%
   \email{Alexander Churbanov\correspondingauthor - alexander@big.ac.cn}
 and
   Levon Abrahamyan\correspondingauthor$^2$%
   \email{Levon Abrahamyan\correspondingauthor - labrahamyan2@unl.edu}%
}



\address{%
    \iid(1)Beijing Institute of Genomics (BIG), Building G, No.7 Beitucheng West Road, Chaoyang District, Beijing 100029, P.R.China\\
    \iid(2)School of Biological Sciences, Nebraska Center for Virology, University of Nebraska-Lincoln, 4240 Fair Street, Morrison Center, Lincoln, NE 68583-0900, USA
}%

\maketitle


\begin{abstract}
\paragraph*{Background:}
Biomedical advances have led to a relaxation of natural selection in the human
population in developed countries. In the absence of strong purifying selection,
spontaneous and frequently deleterious mutations tend to accumulate in the human
genome and gradually increase the genetic load; that is, the frequency of
potentially lethal genes in the gene pool. It is not possible to assess directly
the negative impact of the genetic load on modern society because it is influenced
by many factors such as constantly changing environmental conditions and continuously
improving medical care. However, gradual increase in incidence of many complex
disorders suggests deleterious impact of the genetic load on human well being.
Recent advances in \emph{in vitro} fertilization (IVF) combined with artificial
twinning and improved embryo cryoconservation offer the possibility of preventing
significant accumulation of genetic load and reducing the incidence of hereditary
disorders.
\paragraph*{Discussion:}
Many complex diseases such as type 1 and 2 diabetes, autism, bipolar disorder,
allergies, Alzheimer disease, and some cancers show significantly higher concordance
in monozygotic (MZ) twins than in fraternal twins (dizygotic, DZ) or parent-child
pairs, suggesting their etiology is strongly influenced by genetics. Preventing these
diseases based on genetic data alone is frequently impossible due to the complex
interplay between genetic and environmental factors. We hypothesize that the incidence
of complex diseases could be significantly reduced in the future through a strategy
based on time-separated twinning. This strategy involves the collection and fertilization
of human oocytes followed by several rounds of artificial twinning. If preimplantation
genetic screening (PGS) reports no aneuploidy or known Mendelian disorders, one of the
MZ siblings would be implanted and the remaining embryos cryoconserved. Once the health
of the adult MZ sibling(s) is established, subsequent parenthood with the cryoconserved
twins could substantially lower the incidence of hereditary disorders with Mendelian or
complex etiology.
\paragraph*{Summary:}
The proposed method of artificial twinning has the potential to alleviate suffering and
reduce the negative social impact induced by dysgenic effects associated with known and
unknown genetic factors. Time-separated twinning has the capacity to prevent further
accumulation of the genetic load and to provide source of isogenic embryonic stem cells
for future regenerative therapies.
\end{abstract}
\begin{keywords}
Artificial twinning, \emph{in vitro} fertilization, complex diseases, prevention, preimplantation genetic screening
\end{keywords}


\ifthenelse{\boolean{publ}}{\begin{multicols}{2}}{}

\section*{Background}

Living standards in developed countries have improved to the point where natural selection
is no longer a major driver of human evolution; thus, spontaneous mutations, i.e. {\em de novo}
genetic variants, are routinely passed on to the next generation. Mutations that might have been
selectively eliminated in the past (Figure 1(b)) now accumulate \cite{Crow2000, AustralianAboriginal2001, MichaelLynch2010}
and increase the frequency of potentially lethal genes in the human gene pool, the ``genetic load''.
Mutations will continue to accumulate until a new equilibrium is reached between the relaxed
selective forces of the modern living environment and the higher genetic load. In theory,
mortality before reproductive age is predicted to be 20\% when equilibrium is reattained, the same as for our ancestors
living hundreds of years ago (Figure 1(a)), but in this case medicine would be
working at full capacity and taking resources from productive activity \cite{Muller1950}.

The effectiveness with which modern medicine has reduced mortality and increased life span
creates a false impression that the population has become well adapted to the environment
and that {\em de novo} variants are harmless. However, it has been demonstrated that the majority
of such mutations are either neutral or deleterious
\cite{DrosophilaMutations2007,HaplotypesReconstruction2010,PreviousMutEstimates2000, highDeleteriousRatesHominids1999, selectivelyConstranedHominid2010, lossOfFunction2012} and their gradual accumulation in the genome significantly decreases the chances for human survival without medical intervention.
This leads to the phenomenon of a ``debtor generation'', in which individuals refuse to live and
struggle for existence under primitive conditions in order to attain equilibrium in the mutant
gene frequency under natural purifying selection (see Figures
1(b) and 2(a)). This generation
would transfer mutations to their descendants, which may lead to catastrophic health
consequences in the future \cite{Muller1950}. H.J. Muller estimated that the genetic load accumulated in 8 human generations ($\sim240$ years) living under constantly improving medical services
with zero mortality rate due to natural selective forces is equivalent to that acquired
in a single exposure to $200R$ of gamma radiation 2 kilometers from Ground Zero in Hiroshima.
Michael Lynch in his inaugural article \cite{MichaelLynch2010} estimated fitness decline to be in the range
of 12\% to 60\% in 200 years under the same conditions.

One example of medicine losing its effectiveness is the incidence increase of antibiotic-resistant strains
of {\em Staphylococcus pneumoniae} and {\em S. aureus}, which may be more pathogenic in individuals
with relaxed fitness \cite{AustralianAboriginal2001}. Another example is Shima hospital in Hiroshima,
which was the hypocenter of a nuclear explosion on August 6, 1945 that instantly killed all
on-site medical personnel and thus deprived surviving victims of timely medical assistance.

\begin{figure}[H]
\centering
\includegraphics[width=1.0\linewidth]{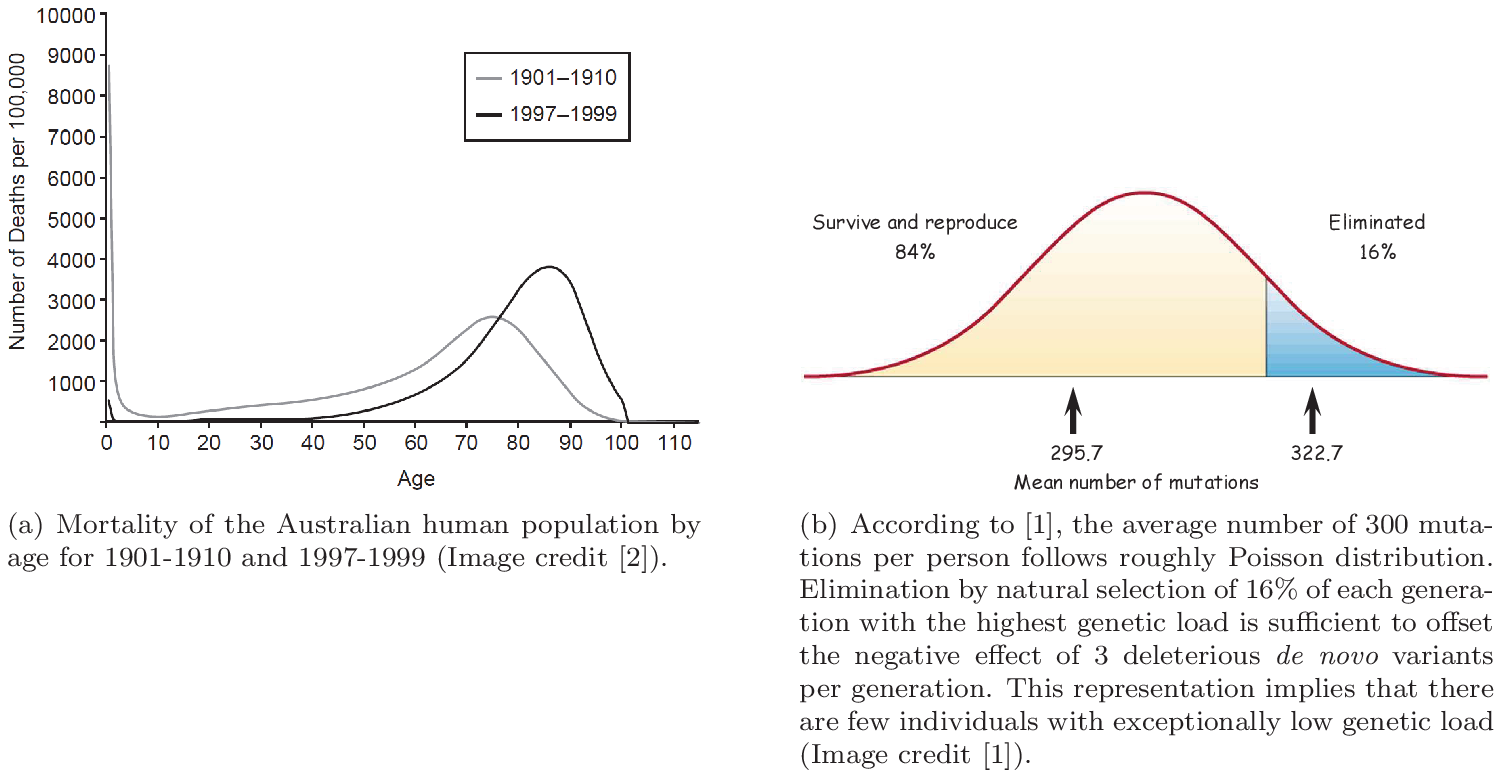}
\caption{Purifying natural selection.}
\label{fig:PurSelWorks}
\end{figure}

Extrapolation of Australian vital statistics \cite{AustralianAboriginal2001} reveals that $\sim 98\%$ of individuals in developed countries survive through their reproductive period (see Figure 1(a)) which supports limited influence of natural selection on reproductive success. Statistics presented in Table \ref{tab:mortalityWHO} indicates that children under 5 mortality in the developed countries is an order of magnitude lower compared to the least developed African region. With majority of individuals surviving till reproductive age sexual selection \cite{CDarwin1871} may play increasing role directing human evolution in developed countries with obscure impact on overall fitness \cite{AustralianAboriginal2001}.

\begin{table}[H]
\centering
\caption{Mortality rates (per 1,000 live births) for 2007 reported by the World Health Organization (WHO) in 2009 \cite{WHOstatistics}.} \label{tab:mortalityWHO}
\begin{tabular}{|c|c|}
  \hline
  Territories & Children under five mortalities\\
  \hline
  African Region & 145 \\
  Eastern Mediterranean Region & 82 \\
  South-East Asia Region & 65 \\
  Western Pacific Region & 22 \\
  Region of the Americas & 19 \\
  European Region & 15 \\
  \hline
\end{tabular}
\end{table}

In the absence of efficient natural selection, eugenics seems to be the only method that would
increase the fitness of humans. Although politically motivated radical interpretations of
negative eugenics led to one of the most tragic chapters in human history \cite{InTheNameOfEugenics1985},
many methods of positive eugenics appear fully compatible with the framework of modern
democratic societies. These include birth control, artificial insemination, {\em in vitro}
fertilization (IVF), prenatal genetic diagnostics such as trisomy \cite{DennisLoTrisomyTesting2011}
and carrier \cite{CarrierTesting2011,mitochondrialDNAtest2012} testing, mitochondrial DNA
swapping \cite{preventingMitochondriaDisease2010}, and artificial human embryo twinning \cite{HumanTwinning2010}.

One strategy to prevent excessive accumulation of genetic load is to borrow genetic material
from the past generation (see Figure 2(b)). The creation of the ``Hermann
J. Muller Repository for Germinal Choice'' sperm bank, which existed in Escondido,
California between 1980 and 1999, was an attempt to implement such a strategy. The cryoconserved
sperm collected from outstanding artists, businessmen, athletes and scientists resulted in the births of 229 children.
Further developments in cryoconservation techniques have allowed entire embryos \cite{EmbryoCryoconservation1983}
to be frozen, which offers previously unconsidered possibilities for reproductive assistance.
One objective of reproductive assistance is to lower the incidence of many hereditary disorders,
which could be accomplished by surrogate parenthood with eggs and sperm taken from unaffected donors.
However, estimating progeny health based on that of egg and sperm donors is inaccurate, because
it depends on many factors beyond our control.

Twin studies have long been recognized as a powerful technique for studying complex phenotypes \cite{ValueOfTwinStudy2012}.
Many complex diseases such as type 1 \cite{TypeGBUSA2001,IDDM1993,significanceT11988,FinnishFollowUp2003,DanishDiabetes1995,DiabetesConcordance2001} and type 2 \cite{earlyStudyDiabeticConcordance1987,DiabetesReviews1998,JapanDiabetes1988,DiabetesConcordanceFinland1992,NCBIdiabetes1999} diabetes, autism \cite{InfantileAutism1977,AutismTwinsEurope1989,BritishTwinStudy1995,AutismTwinStudiesReview2011,AutismVsEnvironment2011}, bipolar disorder \cite{DanishBipolar1977,bipolarFinland2004}, allergy \cite{PeanutAllergyTwins2000,SwedishAllergy1971,OverestimatedAllergy1981,AstmaConcordance2001}, Alzheimer disease \cite{Alzheimers2006}, and some types of cancer \cite{CancerConcordance2000} display significantly higher concordance rates in monozygotic (MZ) twins than in dizygotic (DZ) twins or parent-child pairs, suggesting a large genetic component to their etiology.
Based on this, the use of a time-separated twinning strategy guided by the health status of the adult
siblings could not only prevent further accumulation of genetic load, but also significantly reduce
the incidence of devastating complex diseases. This strategy involves the collection and fertilization
of human oocytes followed by several rounds of artificial twinning (microsurgical splitting of an
embryo at 6 or 8 cell stage). If preimplantation genetic screening
(PGS) reports no aneuploidy or known Mendelian disorders, one of the MZ siblings would be implanted and
the remaining embryos cryoconserved. Once the good health of the adult MZ sibling(s) is established,
subsequent parenthood with the cryoconserved twins could substantially lower the incidence of hereditary
disorders with complex etiology and virtually eradicate simple Mendelian disorders.
The strategy is particularly attractive in the light of recent reports from developed countries
suggesting an alarming increase in the incidence of many diseases that were considered rare only
a few decades ago \cite{YoungerDiabetesIncidence1999,preventingDiabetes2009,CeliacDisease2009,autismPrevalence2009,CDCreport2009,report2008}.

\begin{figure}[H]
\centering
\includegraphics[width=1.0\linewidth]{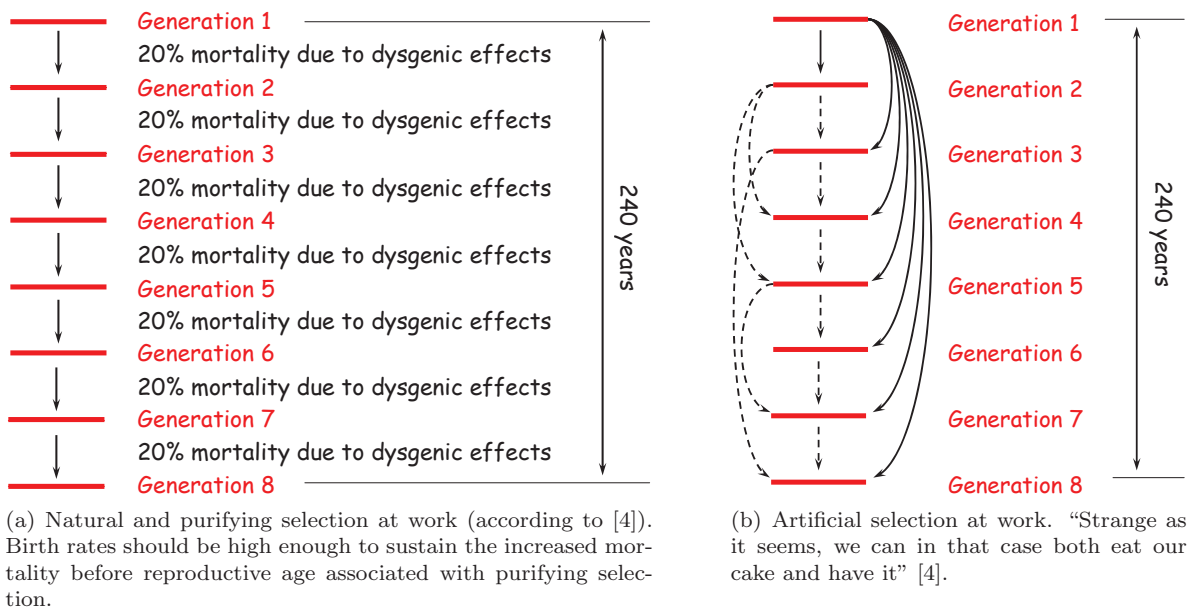}
\caption{Natural and artificial selection.}
\label{fig:NatArtSel}
\end{figure}

In this study, we explore the possibility of using a time-separated twinning strategy to reduce the
incidence of complex non-Mendelian diseases in modern society. We consider the limitations, benefits,
and ethical issues concerning the proposed method.

\section*{Discussion}

\subsection*{Accumulation of genetic load and associated health problems}

Genome sequencing studies have shown that all humans carry many genetic variants predicted to cause loss of function (LoF) of the encoded proteins, suggesting significant redundancy that exists in the human genome. Recently, 2,951 candidate LoF variants were identified from whole-genome sequencing of 185 individuals as part of the 1000 Genomes Project pilot phase \cite{lossOfFunction2012}. Estimates based on the LoF variants predict that the average human genome typically contains $\sim 100$ genuine LoF variants with $\sim 20$ genes completely inactivated.

In addition to deleterious mutations already present in the human genome, approximately 77 {\em de novo}
mutations appear from one generation to the next (a rate of $\sim 1.1\times10^{-8}$ per position per
haploid genome) \cite{HaplotypesReconstruction2010}. Sequencing from sperm cells confirmed presence of
25-36 high confidence candidate point {\em de novo} mutations in each sperm cell \cite{sequencingFromSperm2012}.
Kong {\em et al.} \cite{NewBornMutations2012} showed that genome of a newborn contains $\sim 60$ {\em de novo} variants,
and that this number strongly correlates with the father's age at the time of conception.

It has been established that fathers of advanced age accumulate nucleotide mutations,
while older mothers tend to produce offspring with aneuploidy defects \cite{Crow2000}.
Autism (ASD) has been strongly associated with {\em de novo} mutations \cite{AutismMut2012},
therefore epidemiological studies report significant correlation between paternal age
and risk of ASD \cite{ASDage2007}, schizophrenia \cite{Schizophrenia2001} and bipolar
disorder \cite{AdvancedPaternalAgeBipolar2008,Bipolar2010}. Higher overall mortality
has been reported for children of elder fathers \cite{PaternalAgeChildrenMortality2008}.
These findings provide evidence of deleterious impact of elevated genetic load on human health.

\subsection*{Preventing aneuploidy and Mendelian disorders}

There have been several recent investigations into the diagnosis and prevention
of disorders that result from aneuploidy or single gene mutations. For example,
one study of a patient with a mitochondrial disease examined mitochondrial DNA
(mtDNA) as well as exons of 1,600 nuclear genes involved in mitochondrial biology
or Mendelian disorders with multi-system phenotypes, thereby allowing for
simultaneous evaluation of multiple disease loci \cite{mitochondrialDNAtest2012}.

Scientists at the National Center for Genome Resources (NCGR; New Mexico, USA)
have recently completed a study on preconception carrier screening for 448 severe
recessive childhood diseases. In this study, 7,717 regions from 437 target genes
were enriched by hybrid capture or microdroplet polymerase chain reaction (PCR) and
subjected to next-generation sequencing followed by stringent bioinformatic analysis.
In 104 unrelated DNA samples, the average genomic carrier burden for severe pediatric
recessive mutations was 2.8 and ranged from 0 to 7 \cite{CarrierTesting2011}. The
proposed use of this technology would include preventive carrier testing and prenatal
diagnostics together with post-natal analysis to improve the efficiency of medical
therapy. It is important to note that parental genome testing can only suggest the
possibility of homozygosity on recessive alleles of interest; only fetal DNA analysis
can confirm if the newborn will suffer from a specific Mendelian disorder.

Preimplantation genetic diagnosis (PGD) has already contributed greatly
to the eradication of severe genetic diseases while avoiding selective pregnancy
termination, a significant but often overlooked accomplishment. In majority of PGD cycles
reported to the European Society of Human Reproduction and
Embryology PGD consortium, cleavage-stage biopsy was performed at the 8-cell
stage (third day of development) where one of the blastomeres containing a
nucleus is gently aspirated \cite{Eshre2010}. This means that the genetic contribution of both
parents can be studied for homozygosity on recessive alleles or the presence
of a dominant mutant allele. Single-cell PCR is used to amplify the genetic
material for subsequent next-generation sequencing.

Unless a severe disorder is detected, the embryo is implanted in the uterus
on the fifth day post-fertilization. Successful prenatal screening for aneuploidy
has recently been demonstrated \cite{DennisLoTrisomyTesting2011}. Next-generation
DNA sequencing of blood samples from 753 women with high-risk pregnancies demonstrated
with high accuracy and without false negative results that trisomy 21 (Down syndrome)
was present in the fetuses of 86 women \cite{DennisLoTrisomyTesting2011}.
The recently launched \texttt{GeneInsight}$^{\circledR}$ suite of software (Partners
Healthcare, Cambridge, MA, USA) for diagnosis of Mendelian disorders is already
in use in a number of clinical labs \cite{GeneInsightSuite2011}.

Several promising preventive therapies for genetic disorders have been described
recently, one of which involves replacing damaged mitochondrial DNA (mtDNA swapping)
in fertilized human eggs. mtDNA is maternally inherited and accumulates mutations at
a much higher rate than nuclear DNA. Approximately one in 4,000 children develops a
mitochondrial disease by the age of 10; such diseases are often debilitating, sometimes
fatal, and currently incurable. Researchers at Newcastle University
\cite{preventingMitochondriaDisease2010} successfully replaced the mtDNA in 8 human
embryos, which were then sustained for 6 days, long enough for them to become
blastocysts with about 100 cells. This study used a variation of the mtDNA replacement
technique \cite{MetalipovMtDNAswap2009}, which originally was used with unfertilized
rhesus macaque eggs, rather than with zygotes. Three macaques with swapped mtDNA were
born and developed normally. This successful demonstration of mtDNA swapping opens up
the possibility of other preventive therapies in the future.

In conclusion, recent studies provide substantial evidence that {\em de novo} mutations
with severe disease consequences could be detected easily by either PGD or free-floating
fetal DNA screening. Aneuploidy, mitochondrial diseases, and simple Mendelian disorders
could thus be prevented through routine prenatal genetic counseling. In contrast to these
disorders, prevention of complex diseases based on genetic data alone is a daunting task
because many gene variants are only weakly associated with a disorder. For this reason,
time-separated twinning is a more promising strategy to decrease the incidence of complex
disorders.

\subsection*{Cloning and stem cell therapy}

Since the birth of Dolly the sheep \cite{Dolly1997}, several mammalian species, including goats,
cattle, pigs, mice, rabbits, and cats, have been cloned using somatic cell nuclear transfer (SCNT) \cite{cloningFarmAnimals2003, progressAnimalCloning2002, MetalipovStemCells2009}. Despite the initial success of embryonic blastomere nuclear transfer that led to the birth of 2 unrelated rhesus monkeys \cite{RhMonkeys1997}, efforts to clone rhesus monkeys using SCNT have been unsuccessful \cite{MitalipovClone2002,nuclTransfMonkeys1999,MolCorrelatesNucTransfer2003}. To date, very few blastocysts ($\sim$1\% according to \cite{MitalipovClone2002}) and no pregnancies \cite{MitalipovClone2002,MetalipovStemCells2009} have resulted from SCNT in rhesus monkeys.

Although cloning of adult cattle through SCNT has been demonstrated, it remains an expensive and inefficient technology that is used primarily by the pharmaceutical industry rather than for mainstream agricultural production \cite{AssistedReproductionCows2005}. Consistent with the inefficiency of somatic cell cloning of domestic species and difficulties in obtaining large numbers of oocytes, SCNT would also be a highly inefficient method for cloning of primates. Moreover, even if primate clones were produced by this method one day, they would exhibit various degrees of mitochondrial heterogeneity \cite{NucTransferClonedShips1999} and would not be true clones {\em per se}. The health concordance of such clones could be problematic because mtDNA of the denucleated oocyte may not be compatible with the transferred nucleus. For these reasons, artificial twinning is the only realistic method for primate cloning \cite{MonkeyTwinningStrategy2004}.

One of the challenges of stem cell transplantation is overcoming histoincompatibility between the host and donor cells \cite{SafetyIssuesIncompatible2003,StemCellReplacement2003}.
The dominant alloantigens or antigenic proteins on the surface of transplanted cells are the blood group antigens (ABO) and the major histocompatibility complex (MHC) proteins, which in humans are termed human leukocyte antigens (HLA). The HLA genes are highly polymorphic, there are several thousand known HLA class I alleles and $>1000$ alleles for class II according to \url{http://hla.alleles.org/}. For that reason there may be only one in several million chance of finding a donor-recipient match \cite{numberOfHLAlinesNeeded2005}. Therefore, there is a commonly recognized need to develop efficient approaches for deriving histocompatible pluripotent stem cells.

The conceptual unification of SCNT and ESC derivation technology suggests that it might be possible to produce preimplantation human embryos by SCNT and then derive isogenic ESC from the resulting embryos \cite{FutCloning1999,HumTherapeuticCloning1999}. Feasibility of therapeutic cloning in primates remains illusive because early attempts found that human and nonhuman primate SCNT embryos did not develop into blastocysts and typically arrested at early cleavage stages \cite{MitalipovClone2002,derivationHumanBlastocyst2005}. Researchers in the US who failed to obtain pregnancies from rhesus monkey SCNT embryos, concluded that ``primate NT (nuclear transfer) appears to be challenged by stricter molecular requirements for mitotic spindle assembly than in other mammals'' \cite{primateCloningFailure2003}.

Another method to obtain autologous (the person's own) pluripotent stem cells is by induction \cite{NuclearReprogramming2004}. Despite initially encouraging results in induction of pluripotent stem cells \cite{Definedfactors2007,InductionPluripotency2007,ReprogrammingAdultFibroblastStemCells2007}, recent research indicated that these cells are far more tumorigenic \cite{generationOfMouseInducedPPC2008} than ESCs \cite{MetalipovStemCells2009}. Thus, the use of induced cells in regenerative medicine raises serious safety concerns \cite{GenerationOfGermlineCompetent2007,InducedCellsWorseThanEmbryonic2010} and requires further investigation \cite{tumorogenecityStemCells2009}. Moreover, the epigenome of induced pluripotent stem cells is not authentic and requires further reprogramming \cite{methylationReprogrammingHotspots2011}.

Pluripotent stem cells present in umbilical cord blood of a newborn could be cryoconserved and used later for therapeutic cloning such as for treatment of type 1 diabetes \cite{DiabType1StemCells2008} and cardiovascular repair \cite{cardioRepair2007}. However, without further induction these cells could only be used to treat cardiovascular disorders.

\subsection*{Artificial twinning and efficient cryoconservation}

Artificial twinning strategies have been used for more than 2 decades for breeding of cattle, with the generally positive results \cite{AssistedReproductionCows2005,Lewis1994,conferenceCow1995}. Many thousands of calves have been born worldwide from artificial twinning and there have been no reports of technique-associated abnormalities in offspring, including in Australia where the technique is used extensively \cite{Lewis1994}.
Canadian researchers reported birth of four MZ calves after splitting a single 4-cell stage embryo \cite{FourCalves1995}. Time-separated twinning has already been performed in cows and the growth rates of MZ calves indicated that their developmental patterns were similar despite the different birth dates \cite{cowEmbryoSplitting1991}. This technique is used for predicting the sex and milk producing ability of the offspring and involves progeny testing of a pair of demi-embryos before the decision is made to transfer the second embryo.

The potential application of an artificial twinning technique to humans was suggested by the high twinning success rate for mouse embryos \cite{mouseEmbryo2005}. Blastomere separation and blastocyst bisection have been demonstrated to work consistently well in rhesus monkeys, with 85\%-95\% cases producing MZ twins \cite{MZmonkeyTwinning2002}. In this study, blastocyst bisection was slightly more efficient in the creation of demi-embryos (76/80; 95\%) and in the recovery of pairs (36/40; 90\%). It is interesting to note that the more invasive surgical bisection induced fewer embryo losses than teasing apart the blastomeres at early cleavage stages \cite{MZmonkeyTwinning2002}. Embryo splitting has been proposed to increase the chance of pregnancy in humans and would thus reduce the cost and stress associated with IVF treatment. This technique would also increase the availability of embryos for couples who can not conceive naturally or with IVF \cite{WoodSplitting2001}. For couples with few embryos of good quality available during an IVF cycle, embryo splitting may allow additional embryos to be cryopreserved for subsequent transfer, potentially increasing the likelihood of a pregnancy and even providing time-separated twins \cite{HumanTwinning2010}.

There is evidence that unequal allocation of cells to the twin embryos may lead to phenotypic differences among healthy MZ twins \cite{causesDiscordanceTwinning}, although this does not seem to interfere with normal development of rhesus monkeys \cite{MZmonkeyTwinning2002}. It has been reported that differences in the gene copy number variation (CNV) are the primary causes of phenotypic discordance between MZ twins \cite{CNVtwinsDiff2008}. This mechanism of discordance is similar to the commonly observed mosaicism in healthy somatic cells \cite{somaticMosaicism2011}.
There is also possibility of epigenetic discordance between MZ twins \cite{DiscordanceEpigenetics2002}. Such discordances occur naturally and are taken into account in studies assessing concordance between MZ twins.

In a recent study, human twin blastocysts derived from embryo splitting at the cleavage stage were confirmed to be monozygous by PCR testing of 6 polymorphic single tandem repeat markers within the HLA locus \cite{MZproofTwins2011}. In a study of female MZ twins discordant for multiple sclerosis (MS), no reproducible differences were detected among $\sim 3.6$ million single nucleotide polymorphisms (SNPs) or $\sim 0.2$ million insertion-deletion polymorphisms \cite{MZproofTwins2011}. The siblings of 3 twin pairs were also tested and no reproducible differences were observed in HLA haplotypes, confirmed MS-susceptibility SNPs, CNVs, mRNA and genomic SNP and insertion-deletion genotypes, or the expression of $\sim 19,000$ genes in CD4+ T cells. Of $\sim 2$ million CpG dinucleotides examined, only 2 to 176 differences in methylation were detected between the siblings of the 3 twin pairs \cite{TwinsIdentical2010}.

Not all the embryos obtained by artificial twinning could be immediately implanted, therefore IVF clinics heavily rely on efficient cryoconservation procedures. Cryoloop vitrification gives superior results for survival of rhesus monkey blastocysts \cite{CryoLoopVitrification2001} compared to previous results using controlled-rate cooling. The vitrification freeze-thaw survival rate is $\sim 95\%$ \cite{vitrificationBetter2008,VitrifVsSlowFreezing2009}; therefore, at least 4 embryos must be cryoconserved to achieve a 99.99\% recovery rate. Given that we need to store embryos in 2 different conservation vessels, 8 embryos must be frozen, i.e., there would be only 3 twinnings for each embryo (Level 4 in Figure \ref{fig:twining}) which is only 7 mitotic divisions for each cryoconserved totipotent cell. More twinnings might be necessary to increase the chance of pregnancy with thawed embryos from 43.7\% \cite{FreezeThawSucess2010,vitrifFertility2011} to 100\% \cite{WoodSplitting2001}. The efficiency of embryo twinning is approximately 95\% \cite{MZmonkeyTwinning2002} or less \cite{tetra2000}, suggesting that additional twinnings might be necessary beyond level 4, which could be problematic.

Although under optimal conditions pluripotent ESCs could propagate indefinitely, both in murines \cite{StemCellsIndefinite1981} and humans \cite{Amit2000,Eiges2001,OdoricoES2001}, the number of serial artificial twinning cycles in mammals is normally limited to 3 \cite{SeriallySplitMouse2006}.
Progressively poor results in serial embryo twinnings could be explained by the fact that totipotent cells contained in morula utilize only the maternal mRNA and proteins contained in oocyte \cite{TipOfIceberg2011}. The maternally derived factors are necessary to initiate massive epigenome reprogramming to activate expression of embryonic genes \cite{geneSilencingEmbryo2006,mouseEpigenomeReprogramming2011,TipOfIceberg2011}. Therefore, excessive twinning may deprive the blastomeres of needed maternal nutrients to develop in competent self-sustained blastocysts. Further research is necessary to investigate if twinning competence could be sustained by replenishing maternal factors and by signalling with proper proteins.

The birth of Tetra (Rhesus macaque), a healthy female cloned from a quarter of an embryo, suggests
that at least four healthy MZ siblings could be obtained from a single primate embryo splitting \cite{tetra2000}.
Clonal propagation in Rhesus macaque embryo frequently arrested when splitting beyond sextuplets \cite{tetra2000}.
While propagation beyond quadruplets may need additional research, blastocysts from quintuplets to septuplets could be used for establishing ESC lines
\cite{tetra2000}. Study \cite{fourCellsSplit2008} reported putting 4 human blastomeres from the same 4-cell stage embryo individually into empty zona pellucida (ZP) that later developed in blastocysts, confirming total potency of all 4 cells. Contrary to this study, only 2 human blastomeres
out of 16 developed into blastocysts when cultured without ZP, suggesting importance of maternally derived
factors in embryo development \cite{fourCellsSplit2009}. Higher survival rate of individual mouse blastomeres cultured in ZP, compared
to blastomeres cultured without ZP, has been reported \cite{mouseZP2006}.

\begin{figure}[H]
\centering
\includegraphics[width=1.0\linewidth]{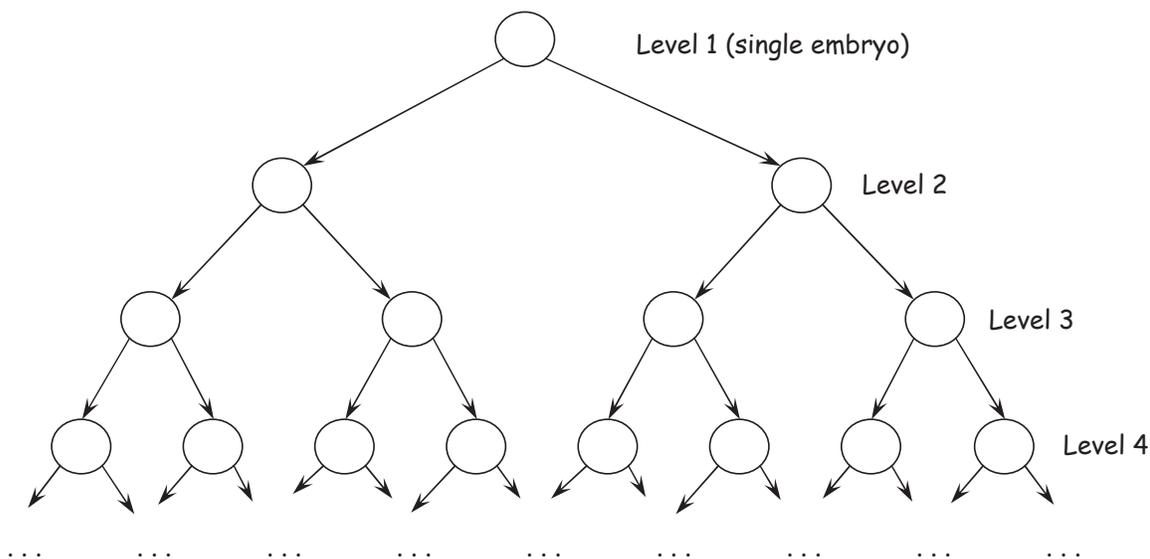}
\caption{Three mitotic divisions are required for the level 1 embryo
         before it becomes competent for twinning at the 6-cell stage,
         in addition to 2 mitotic divisions at each level. The number of
         embryos at each level $L$ is $2^{L - 1}$.}
\label{fig:twining}
\end{figure}

Radiation-induced damage of DNA is the primary concern for ultra-long storage times for
cryopreserved biological samples. A study involving 11,768 cryopreserved human embryos reported that the duration of storage
had no significant effect on post-thaw survival for IVF or oocyte donation cycles or for embryos
frozen at the pronuclear or cleavage stages \cite{CryoconservationNoProblems2008}.
Whole, fertile plants of {\em Silene stenophylla} Ledeb.
(Caryophyllaceae) have been uniquely regenerated from maternal, immature fruit tissue
found in permafrost of late Pleistocene age (31,500-32,100 years old) using {\em in vitro} tissue
culture and clonal micropropagation \cite{permafrostPlant2012}. The total $\gamma$-radiation dose accumulated by the
fruits was 0.07 kGy; the maximal reported dose after which tissues of a complex organism remain viable \cite{permafrostPlant2012}.

A viable strain of {\em Carnobacterium pleistocenium} bacteria $\sim 32,000$ years old has been
successfully isolated from a lenticular ice lens associated with a Pleistocene thermokarst
pond \cite{bacteriaPermafrost2005}. Analyses of 5 samples of ice ranging in age from 100,000
to 8 million years established a bacterial DNA degradative half-life of $\sim 1.1$ million
years (the DNA size is degraded $\sim 50\%$ every 1.1 million years) for the polar regions
that are subject to higher cosmic radiation compared to the rest of the planet \cite{longestBacteriaSurvive2007}. Thus, the
viability of microorganisms found in ice up to $\sim 300,000$ years old could be reliably
re-established \cite{longestBacteriaSurvive2007}.


\subsection*{Modeling the possible prevention of complex disorders with a time-separated twinning strategy}

The accumulation of mildly deleterious missense mutations in individual human
genomes has been proposed to contribute to the genetic basis for complex diseases,
which likely result from a combination of genetic, lifestyle, and environmental factors.

In a recent genome-wide association study of rheumatoid arthritis, thousands of SNPs were
found to explain 20\% of disease risk, in addition to the known associated loci \cite{NatureGeneticAdvance2012}.
Analysis of datasets for 3 additional diseases yielded comparable estimates for celiac disease (43\% excluding the MHC), myocardial infarction and coronary artery disease (48\%), and type 2 diabetes (49\%). These results suggest that complex
disease risk factors are associated with common causal variants located within hundreds of loci and with multiple rare causal variants located in a smaller number of loci \cite{NatureGeneticAdvance2012}. Predictive capacity of clinically significant risk for
different complex diseases based on personal genome sequencing was recently characterized as very
limited \cite{predictivePower2012,23andMeStudy2012}.

Geneticists and epidemiologists often observe that certain hereditary disorders co-occur
in individual patients significantly more (or significantly less) frequently than expected,
suggesting there is genetic variation that predisposes its bearer to multiple disorders,
or that protects against some disorders while predisposing to others \cite{probingGeneticOverlap2007}.

Here, we model the possible prevention of complex disorders with a time-separated twinning
strategy in pure form based on previously established pairwise concordance between MZ twins.
Assuming unchanged environmental conditions, there is no reason to believe that creating
population $y$ by direct replication of founder population $x$ through time-separated twinning
(as shown in Figure \ref{fig:Model}) would increase or decrease disease incidence. We also
hypothesize that time-separated MZ twins would have the same pairwise concordance as the
same-age MZ twins. In reality, concordance between time-separated MZ twins could be lower
due to unshared environmental factors. Following these assumptions we can calculate the
probability of an individual becoming sick in the replicate population $y$ following the Bayesian rule:

\begin{equation}\label{eq:BayesianRule}
    p(sick_y) = p(sick_y|\neg sick_x) \, (1 - p(sick_x)) + p(sick_y|sick_x) \, p(sick_x)
\end{equation}
where $p(sick_y|\neg sick_x)$ is the likelihood that a healthy individual in population $x$ would become sick
in the replicate population $y$, $p(sick_x)$ is the disease prevalence in the founder population $x$, $p(sick_y)$
is the disease prevalence in the replicate population $y$, and $p(sick_y|sick_x)$ is the pairwise concordance, i.e.
the likelihood that the second MZ co-twin would have a disease if his sibling has the condition. According to the
previously made assumptions, $p(sick_x)=p(sick_y)$.

From the biomedical literature on twin concordance we can obtain a general population prevalence of a disease $p(sick_x)$ and pairwise concordance between MZ twins $p(sick_y|sick_x)$. Having this information we can use formula (\ref{eq:BayesianRuleSick}) to estimate the likelihood of an individual becoming sick in the replicate population $y$ given that his MZ co-twin in the founder population $x$ is healthy:

\begin{equation}\label{eq:BayesianRuleSick}
  p(sick_y|\neg sick_x) = \frac{p(sick_y) - p(sick_x) \, p(sick_y|sick_x)}{1 - p(sick_x)} = \frac{p(sick_x)\,(1 - p(sick_y|sick_x))}{1 - p(sick_x)}.
\end{equation}

The population in generation $z$ consists only of MZ twins whose siblings in generation $y$ were healthy, as shown in Figure \ref{fig:Model}, thus substantially reducing the disease incidence in that population, calculated with formula (\ref{eq:BayesianRuleSickBetter})

\begin{equation}\label{eq:BayesianRuleSickBetter}
  p(sick_z) = p(sick_z|\neg sick_y) \, p(\neg sick_y) = p(sick_z|\neg sick_y) \times 1
\end{equation}

given that $p(sick_z|\neg sick_y) = p(sick_y|\neg sick_x)$.

\begin{figure}[H]
\centering
\includegraphics[width=0.8\linewidth]{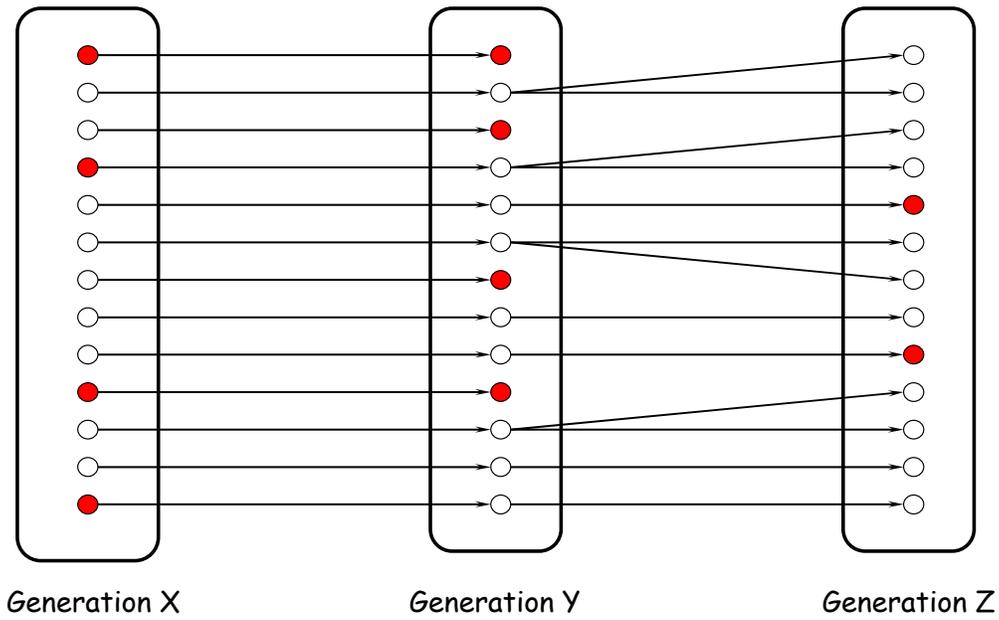}
\caption{A simple model to explain our calculations. Red circles represent
         individuals affected by a certain disease and open circles represent
         healthy individuals in a population.}
\label{fig:Model}
\end{figure}

Based on previously published pairwise concordance rates between MZ twins, we can use the belated
twinning strategy in pure form and the model discussed, at least in theory, to significantly reduce
the incidence of various diseases, as shown in Table \ref{tab:prevention}.

\begin{table}[H]
     \caption{Disease incidence and theoretically possible prevention odds.} \label{tab:prevention}
     \begin{tabular}{|c|l|c|c|c|}
        \hline
        \multirow{2}{*}{\textbf{Study}} & \multirow{2}{*}{\textbf{Type}} & \multirow{2}{*}{\textbf{Disease incidence}} &
        \textbf{MZ twins pairwise} & \multirow{2}{*}{\textbf{Prevention odds}} \\
        & & & \textbf{concordance} & \\
        \hline
        \multicolumn{5}{|c|}{\textbf{Diabetes type 2}} \\
        \hline
        \cite{NCBIdiabetes1999} & & \multirow{4}{*}{7.67\%-13.5\%} & 76\% & 3.75\\
        \cite{JapanDiabetes1988} & & & 83\% & 5.29 \\
        \cite{DiabetesConcordanceFinland1992} & & & 20\% & 1.12\\
        \cite{earlyStudyDiabeticConcordance1987} & & & 85.3\% & 6.12\\
        \hline
        \multicolumn{5}{|c|}{\textbf{Diabetes type 1}} \\
        \hline
        \cite{JapanDiabetes1988} & &\multirow{6}{*}{0.01\% - 0.034\%} & 45\% & 1.82\\
        \cite{DiabetesConcordanceFinland1992} & & & 13\% & 1.15\\
        \cite{FinnishFollowUp2003} & & & 27.3\% & 1.37 \\
        \cite{IDDM1993,significanceT11988,TypeGBUSA2001,FinnishFollowUp2003} & $\leq 10$ years old & & 50\% & 2.0\\
        \cite{FinnishFollowUp2003} & $> 10$ years old & & 16.7\% & 1.20\\
        \cite{DanishDiabetes1995} & & & 38\% & 1.61 \\
        \hline
        \multicolumn{5}{|c|}{\textbf{Cancer}} \\
        \hline
        \multirow{3}{*}{\cite{CancerConcordance2000}} & Breast & 1.92\% & 14\% & 1.14\\
          & Colorectum & \multirow{1}{*}{1.55\%} & 16\% & 1.17\\
          & Prostate & \multirow{1}{*}{1.12\%} & 21\% & 1.25\\
        \hline
        \multicolumn{5}{|c|}{\textbf{Autism spectrum disorder}} \\
        \hline
        \cite{BritishTwinStudy1995} & & \multirow{4}{*}{0.1\%-1\%} & 92\% & 12.37 \\
        \cite{AutismTwinsEurope1989} & & & 91\% & 11.0 \\
        \cite{Ritvo1985} & & & 95.7\% & 23.08\\
        \cite{InfantileAutism1977} & & & 82\% & 5.5\\
        \hline
        \multicolumn{5}{|c|}{\textbf{Allergies}} \\
        \hline
        \cite{PeanutAllergyTwins2000} & Peanut & 0.4-0.6\% & 64.3\% & 2.79 \\
        \hline
        \multirow{11}{*}{\cite{AstmaConcordance2001}} & Astma past year $(<50\sim\geq 50)$ yo & 8\%$\sim$3\% & 29\%$\sim$0\% & 1.30$\sim$0.97 \\
        & Hay fever $(<50\sim\geq 50)$ yo & 30\%$\sim$27\% & 39\%$\sim$30\% & 1.16$\sim$1.05 \\
        & Seasonal	 & \multirow{2}{*}{15\%$\sim$11\%} & \multirow{2}{*}{31\%$\sim$18\%} & \multirow{2}{*}{1.23$\sim$1.09} \\
        & rhinoconjunctivitis $(<50\sim\geq 50)$ yo & & & \\
        & Eczema $(<50\sim\geq 50)$ yo & 24\%$\sim$16\% & 34\%$\sim$30\% & 1.16$\sim$1.20 \\
        & Pets $(<50\sim\geq 50)$ yo & 13\%$\sim$4\% & 39\%$\sim$7\% & 1.43$\sim$1.03 \\
        & Pollen $(<50\sim\geq 50)$ yo & 21\%$\sim$14\% & 32\%$\sim$17\% & 1.17$\sim$1.03 \\
        & Dust $(<50\sim\geq 50)$ yo & 22\%$\sim$12\% & 43\%$\sim$8\% & 1.37$\sim$0.95 \\
        & Insect bites $(<50\sim\geq 50)$ yo & 10\%$\sim$11\% & 20\%$\sim$11\% & 1.12$\sim$1.00 \\
        & Cat IgE+ $(<50\sim\geq 50)$ yo & 12\%$\sim$4\% & 28\%$\sim$44\% & 1.22$\sim$1.74 \\
        & Grass IgE+ $(<50\sim\geq 50)$ yo & 21\%$\sim$10\% & 56\%$\sim$35\% & 1.77$\sim$1.38 \\
        & Der p 1 IgE+ $(<50\sim\geq 50)$ yo & 22\%$\sim$9\% & 54\%$\sim$14\% & 1.69$\sim$1.06 \\
        \hline
        \cite{ItalyGUT2006} & Celiac disease & 0.75\% & 71.4\% & 3.47 \\
        \hline
     \end{tabular}
  \end{table}

\subsubsection*{Type 1 Diabetes}

According to NIH MedlinePlus (\url{http://www.nlm.nih.gov/medlineplus/diabetestype1.html}), type 1 diabetes (T1D) results from autoimmune destruction of the insulin-producing $\beta$-cells of the pancreas. The subsequent lack of insulin leads to increased glucose levels in the blood and its excretion in urine. T1D is fatal unless treated with regular injections of insulin, guided by careful monitoring of blood glucose levels.

Progression to clinical T1D typically requires the unfortunate combination of genetic disease susceptibility, a diabetogenic trigger, and high exposure to a driving antigen \cite{EnvType1_2005}. Several environmental factors have been proposed to trigger T1D but causative roles have not yet been established \cite{EnvTriggersDiab2001,noEnvironmentDiabetes1992,vitaminDintake2001}.

The exact molecular mechanisms leading to the development of T1D remain elusive, despite years of research. T1D is a polygenic disease, meaning many different genes contribute to its onset. The most important disease susceptibility genes are located in the HLA class II locus on the short arm of chromosome 6 \cite{EstimatingRiskType1:2002,HLAdiabetesT1}. Genes in the HLA region substantially influence the risk of T1D and familial clustering \cite{HighDiabetesRiskFinland1988,HLAantigensInsulin1991}. Nevertheless, only a relatively small proportion, $\sim10\%$, of genetically susceptible individuals progress to clinical disease \cite{EnvType1_2005}. The high-risk Hph I insulin genotype increases the likelihood of identical twins being concordant for T1D and the ``load'' of both MHC and non-MHC diabetes susceptibility genes increases T1D predisposition \cite{DiabetesConcordance2001}. Activating mutations in the KCNJ11 gene encoding for the Kir6.2 subunit of the $\beta$-cell ATP-sensitive potassium channel have been shown to be a common cause of permanent neonatal diabetes \cite{DiabeticMosaics2004}. The possibility of germline mosaicism should be considered when counseling recurrence risks for the parents of a child with an apparently de novo KCNJ11 activating mutation \cite{DiabeticMosaics2004}.

The age-adjusted incidence of T1D varies from $<0.1/100,000$ per year in China and
Venezuela to $>36/100,000$ per year in Sardinia and Finland, which represents a more
than 350-fold variation \cite{worldwideDiabetesDistribution2000}. In most populations,
the T1D incidence increased with age and was the highest among children 10-14
years of age \cite{worldwideDiabetesDistribution2000}.

Previous analyses showed that Finland's record incidence of T1D increased predominantly in younger age groups \cite{FinnishFollowUp2003,IncreasedDiabetesFinns1999,YoungerDiabetesIncidence1999}. In children aged 1-4 years,
the increase was 4.2\% per year, and the overall 32-year relative increase was 338\% \cite{YoungerDiabetesIncidence1999}.
Statistical analysis of the incidence of T1D in Hungary in the 20-year period between 1978 and 1997
showed a steady increase with an average annual rate of 4.8\% \cite{HungaryDiabetes1999}. An annual increase
of 4.4\% in the incidence of T1D in Hungarian children aged 0-14 years was reported for 1989-2009 \cite{IncreaseDiabetesHungary2011}.

A concordance of $\sim50\%$ in T1D has been reported between MZ twins less than 10 years of age \cite{IDDM1993, significanceT11988, TypeGBUSA2001, FinnishFollowUp2003}. The higher concordance for patients diagnosed at younger ages suggests an elevated contribution of the genetic component in recent cases of the disease.

\subsubsection*{Type 2 diabetes}

According to Wikipedia (\url{http://en.wikipedia.org/wiki/Diabetes_mellitus_type_2}), type 2 diabetes (T2D) is ``a metabolic disorder that is characterized by high blood glucose in the context of insulin resistance and relative insulin deficiency''. T2D is ``typically a chronic disease, associated with a ten year shorter life expectancy''. Long-term complications from high blood glucose can include heart attack, stroke, diabetic retinopathy, kidney failure requiring dialysis, and poor peripheral circulation leading to limb amputation.

Before the agricultural advent members of the hunters-gathers society had a very unpredictable diets.
In these conditions individuals with increased ability to turn food into fat had a survival advantage \cite{DiabetesEnvironment2005}.
However, in the modern world obese people are more prone to T2D and the associated health complications.
A previous study called the China Da Qing Diabetes Prevention Study reported a 51\% reduction in new cases
of diabetes among Chinese patients as a result of lifestyle intervention \cite{diabetesInterventinChina2008}.

According to \cite{preventingDiabetes2009} nearly 200 million people worldwide have T2D, and this number is predicted to increase to 333 million by 2025. In China, $\sim 23.46$ million people currently have T2D, and this number is projected to increase to $\sim 42.30$ million by 2030. Between 1996 and 2006, the prevalence of T2D mellitus increased rapidly in urban China, from 4.58\% to 7.67\%, and was much higher in major cities (6.1\%) than in small cities (3.7\%) and rural areas (1.8\%). The annual per patient direct medical costs of healthcare associated with T2D were estimated to be \$1,798 with complications, compared with \$484 for those without complications. Based on case numbers in 2007 and projected case numbers in 2030, the direct medical costs of T2D and its complications were estimated to rise from \$26.0 billion in 2007 to \$47.2 billion in 2030 \cite{preventingDiabetes2009}.

Higher incidence of T2D in Japan is associated with increased longevity and lifestyle changes \cite{DiabetesGrowthInJapan2009}. Approximately 13.5\% of the Japanese population now has either T2D or impaired glucose tolerance. This high prevalence is associated with a significant economic burden, with T2D accounting for up to 6\% of the total healthcare budget. The costs of T2D are increased in patients with co-morbidities such as hypertension and hyperlipidemia and in patients who develop complications, of which retinopathy has the highest cost \cite{DiabetesGrowthInJapan2009}.

T2D is a polygenic disease with a strong genetic basis \cite{DiabetesReviews1998}; in most cases of the disease each gene makes a small contribution to an increased probability of disease development. As of 2011, more than 36 genes are known to contribute to the risk of T2D \cite{PredispositionType2Diabetes2011}. Four variants near the $HNF4\alpha$ gene were identified that occur more frequently in people with T2D than in those without the disease \cite{dia2Succ2004,HNFtf2004,HNFsci2004}. However, all of these genes collectively account for only 10\% of the total genetic component of the disease \cite{PredispositionType2Diabetes2011}. Offspring of patients with T2D have a 40\%-60\% chance of developing the disease and an increased frequency of impaired glucose tolerance (IGT) \cite{familialClusteringDiabetesComplications2000}.

Concordance for T2D among 250 MZ born between 1917 and 1927 was found at the level to be 58\%, and 65\% of non-diabetic MZ co-twins of diabetic twins had elevated glucose values \cite{earlyStudyDiabeticConcordance1987}. If the elevated glucose levels were also considered, the concordance in this study was 85.3\% \cite{earlyStudyDiabeticConcordance1987}. In another study, T2D concordance between MZ twins was 83\% and between DZ twins was 40\% (4/10)
and the concordance was found significantly greater in T2D than in T1D \cite{JapanDiabetes1988}. In concordant pairs the presence or
absence of various complications agreed in 68\%-97\% \cite{JapanDiabetes1988}. The study of Medici et al. \cite{NCBIdiabetes1999} examined twin pairs that were part of the British Diabetic Twin Study between May 1968 and January 1998. The observed rates of concordance for T2D at 1, 5, 10, and 15 years follow-up were 17\%, 33\%, 57\%, and 76\%, respectively. The concordance rate for any abnormality of glucose metabolism (either T2D or IGT) at 15 years follow-up was 96\%. The concordance rate for T2D in MZ twins was very high, even in twins initially determined to be discordant for T2D.

Unusually low probandwise and pairwise concordance rates for T2D were reported among Finnish MZ twins (34\% and 20\%, respectively), and DZ twins (16\% and 9\%, respectively) \cite{DiabetesConcordanceFinland1992}. Contrary to other reports, this study found greater heritability for T1D than for T2D, where both genetic and environmental effects seemed to play a significant role \cite{DiabetesConcordanceFinland1992}. In a follow-up of this study \cite{FinnishFollowUp2003}, the pairwise concordance among MZ for T1D was elevated from the initial estimate of 13\% \cite{DiabetesConcordanceFinland1992} to 27.3\%; however, no updates were provided for T2D concordance.

\subsubsection*{Celiac disease}

Celiac disease (CD) is an inherited autoimmune disorder that causes the body's immune system to attack the small intestine, according to the US National Institutes of Health and the University of Chicago Celiac Disease Center. Nearly 5 times as many people have CD today than during the 1950s, according to one recent study \cite{CeliacDisease2009}. Level of tissue transglutaminase and, if abnormal, for endomysial antibodies were compared between blood sera collected (during 1948-1954) from 9,133 healthy young adults at Warren Air Force Base and sera recently collected from 7,210 gender and age at sampling matched individuals from Olmsted County, Minnesota. The study found that the rate of CD has doubled every 15 years since 1974, and is now believed to affect one in every 133 US residents. Undiagnosed CD was associated with a nearly 4-fold increased risk of death \cite{CeliacDisease2009}. Pairwise concordance reported in \cite{ItalyGUT2006} was significantly higher in MZ (71.4\%) than in DZ (9.1\%) twins providing evidence for a very strong genetic component in multifactorial celiac disease.

\subsubsection*{Cancer}

Cancer is a broad group of diseases, all involving unregulated cell growth. According to American Cancer Society (\url{http://www.cancer.org}), occupational hazards and poor lifestyle are among the primary risk factors for the disease. Cancer also has some susceptibility gene variants, among these are mutations in TP53, BRCA1 and BRCA2 genes (population carrier frequency $\leq0.1\%$), associated with the breast cancer \cite{mutationsLandscape2008}; DNA mismatch-repair genes, associated with the hereditary nonpolyposis colorectal cancer; and the candidate gene HPC1, associated with prostate cancer \cite{CancerConcordance2000}. However, the incidence of these mutations is too low to explain more than a small fraction of the genetic predisposition found in a concordance study of Scandinavian twins \cite{CancerConcordance2000}.

The very high incidence in MZ twins of patients indicates that a high proportion, and perhaps the majority, of breast cancers arise in a susceptible minority of women \cite{HighBCincidence2000}. It is known that the breast cancer predisposition is transmitted as an autosomal dominant trait in families harboring mutations \cite{mutationsLandscape2008}. Statistically significant effects of heritable risk factors were found for prostate cancer (42\%), colorectal cancer (35\%), and breast cancer (27\%), which suggests a minor contribution of inherited genetic factors to susceptibility to most types of neoplasms \cite{CancerConcordance2000}.

\subsubsection*{Autism}

According to Autism Speaks (\url{http://www.autismspeaks.org/}), ASD is a complex
developmental disability that causes problems with social interaction and lack of reciprocity.
Symptoms usually start before age 3 and can cause delays or problems in many different skills
that develop from infancy to adulthood.

ASD appears to involve multiple genes each with its own risk factor \cite{AutismPredictors1999}.
Possible susceptibility regions include chromosome regions 1p, 2q, 7q, 13q, 16p, and 19q \cite{autismScreen1999}.
A two-hit model has been proposed for ASD in which a 16p12.1 microdeletion both predisposes to
neuropsychiatric phenotypes as a single event and exacerbates neurodevelopmental phenotypes in
association with other large deletions or duplications \cite{autismMicrodeletion2010}.

Several decades ago ASD has been considered as a rare behavioral disorder affecting children of cold, emotionally distant intellectual mothers. Leo Kanner in his influential paper of 1943 \cite{autisticDisturbances1943} called attention to what he saw as a lack of parental warmth and attachment to their autistic children. In his 1949 paper \cite{AutismDynamics1949}, he attributed autism to a ``genuine lack of maternal warmth'' and blamed autism on bad parenting and cold, withdrawn ``refrigerator mothers''. It appears that Leo Kanner was confusing cause and effect since he accused only parents for the genuine lack of attachment with their autistic kids. In his study of 1943 he consistently ignored the fact of normal reciprocity between the same parents and other unaffected siblings. During his interview of 1960 to the Time Magazine he characterized the mothers of autistic children as ``just happening to defrost enough to produce a child''.

ASD was not appreciated as a mainly genetic disorder until a 1977 study \cite{InfantileAutism1977}
that found pairwise concordance between MZ twins as high as 82\% and only 10\% between DZ twins.
A similarly high concordance of 91\% was reported between Scandinavian MZ twins \cite{AutismTwinsEurope1989}
and 92\% between British MZ twins \cite{BritishTwinStudy1995}. Heritability of autism of over 90\%
was reported in \cite{BroaderAutismPhenotype1997} with no convincing evidence for perinatal factors
playing important roles. The process of twinning itself has not been found as a substantial risk
factor in the etiology autism \cite{AutismTwinning20002}.

The incidence of ASD among US children has reached a staggering 1\% (110 out of 10,000) \cite{autismPrevalence2009,CDCreport2009}. Among Hispanic children, the prevalence of ASD almost tripled from 2.7 per 1000 in 2000 to 7.9 per 1000 in 2006 \cite{hispanicsAutism2012}. The rise in incidence of ASD has been partially attributed to better diagnostics \cite{AustralianASDeffectDiagnostics2009}. The observed prevalence of autism in young children in Denmark was inflated due to shifts in the age at diagnosis in the more recent cohorts compared with the oldest cohort \cite{AutismDenmark2008}. This study supports the argument that the apparent increase in autism in recent years is at least partly attributable to decreases in the age at diagnosis over time.

A recent controversial study on concordance rates in twins with autism \cite{AutismVsEnvironment2011}
suggested that environment, such as conditions in the womb, age of parents, and other factors may play
significant roles. A mathematical model presented in the study estimated environmental influence in the
range 9\%-81\% percent and genetic contribution in the range 8\%-84\%, far too broad to make any definite
conclusions on disease etiology. Of 1,156 pairs of twins that fit the criteria for the study, the
researchers performed further assessments on only 202 pairs. Potential biases in the participation rate
might also be a problem because families with 2 affected siblings might be more willing to participate
thus inflating the environmental contribution. Therefore, accepting the calculated risks reported in
\cite{AutismVsEnvironment2011} for ASD of 38\% coming from genes and 58\% coming from the environment
will bring us back to the pre-1977 era of ``refrigerator mothers''.

Recent study based on whole-exome sequencing of 928 individuals, including 200 phenotypically discordant sibling pairs,
revealed number of highly disruptive {\em de novo} variants affecting brain-expressed genes.
Among a total of 279 identified {\em de novo} coding mutations, two independent nonsense variants
were found to disrupt SCN2A gene, an event highly unlikely by chance \cite{AutismMut2012}.
Trio study that involved sequencing of 677 individual exomes from 209 families of ASD affected child revealed that
39\% (49 of 126) of the most disruptive {\em de novo} mutations map to a
highly interconnected $\beta$-catenin/chromatin remodelling protein network that contains
many genes previously linked to ASD \cite{AutismEichler2012}. Combined with copy number variant
(CNV) data, these results indicate extreme locus  heterogeneity for ASD \cite{AutismEichler2012}.
Another exome sequencing of 175 trios (parents and their affected child) found important but limited
role for {\em de novo} point mutations in ASD, similar to that documented for {\em de novo} copy number
variants \cite{AutismWithCook2012}. This result supports polygenic etiology of ASD in which spontaneous
{\em de novo} mutations compromising integrity of a large number of genes increases risk by 5- to
20-fold \cite{AutismWithCook2012}.

Advanced parental age has been cited as one cause of ASD \cite{AutismParentalAge2008,advancedPaternalAgeAutism2010}.
This observation is in line with the reportedly higher accumulation of {\em de novo} mutations in the sperm of elderly
fathers \cite{Crow2000}. It has been recently shown that 80\% of {\em de novo} point mutations are of paternal origin
that correlate positively with an advanced paternal age \cite{AutismEichler2012,NewBornMutations2012}. Some of these mutations have been
predicted as likely ASD predisposition candidates \cite{deNovoMutations2011,AutismWithCook2012,AutismEichler2012,AutismMut2012}.

\subsubsection*{Allergies}

Allergy occurs when a person's immune system reacts to normally harmless substances in the environment called allergens. Parents with allergies appear to have a much greater chance of having children with allergies \cite{RiskFactorsForAllergy1999} and their children are likely to have more severe allergies than are children of parents without allergies. The likelihood of developing allergies in general, but not a specific allergy, seems to be inherited \cite{RiskFactorsForAllergy1999}. Asthma and other allergic diseases have become much more common in the US in the last 40 years. These diseases affect 40 to 50 million people, more than 20\% of the population \cite{report2008}. The incidence of asthma alone has at least tripled over the past 25 years and affects more than 22 million people \cite{report2008}. These statistics appear even more disturbing considering that between 1990 and 1998 the number of US physicians training in allergy and immunology fellowships declined by 34\% \cite{report2011,techReport41}.

According to Peanut Allergy Online Resource Guide (\url{http://www.peanutallergy.com/}), peanut allergy has a prevalence of 0.4\%-0.6\%.
The disease usually requires careful avoidance of food containing whole peanuts, peanuts particles, peanut dust
cross-contaminating many non-peanut foods, peanut butter and peanut oil. Peanut allergy has been reported to have a
pairwise concordance of 64.3\% between MZ twins and 6.8\% between DZ twins \cite{PeanutAllergyTwins2000}. The higher
concordance rate among MZ twins strongly suggests that there is a significant genetic influence on this allergy.

Early studies of concordance of allergic disease among MZ and DZ twins were based on small clinical case series and
considered asthma, allergic rhinitis, and eczema in combination. Pairwise concordances reported in \cite{OverestimatedAllergy1981}
were 58\% for MZ twins and 38\% for DZ twins. In contrast, a population-based sample of 6,736 Swedish twins \cite{SwedishAllergy1971}
found pairwise concordances of 25\% for MZ twins and 16\% for DZ twins for self-reported history of asthma, hay fever, or eczema.
This difference probably reflects the recognized tendency of clinical case series to overestimate concordance through selective identification of concordant pairs \cite{TendencyToOverestimate1983}. An extensive study on allergies among British MZ and DZ female twins \cite{AstmaConcordance2001} showed generally higher concordance among the cohort younger than 50 years of age. This higher concordance between younger MZ twins might indicate the presence of deleterious {\em de novo} variants compromising the immune system. Accumulation of such variants may also partially explain the substantial growth of allergies observed in recent years \cite{report2008}.

\section*{Conclusions}

Advances in modern medicine in developed countries have considerably reduced mortality among the younger generation. However, such luxury should be considered a gift from our ancestors. H. J. Muller \cite{Muller1950} has mentioned that ``This situation is however due to the fact, so fortunate for all of us in this generation, that our germ plasm was selected, in our more primitively living ancestors, for a world without central heating or refrigerators, without labor-saving mechanisms in the home, in industry or in agriculture, without sewers or bathrooms, and without knowledge of contraceptives, asepsis, antibiotics, calories, vitamins, hormones, surgery or psychosomatic treatment.''

The unfortunate reality is that, despite sanitation and ecological improvements, the genetic load has a tendency to accumulate and without intervention the situation will only worsen. Novel chemistries used in consumer products, radio wave emitting devices and areas contaminated by radioactive fallout create new environment with potentially higher mutagenesis. For future generations this will result in a constant struggle with the hereditary disorders induced by inherited and {\em de novo} variants that will significantly compromise their quality of life bringing it to below the modest standards of living enjoyed by an average healthy person in the 1970s.

Unfortunately, the measures of artificial selection proposed by H. J. Muller \cite{Muller1950} 62 years ago have not led to a measurable reduction in the incidence of human disease. His call for voluntary abstention from reproduction, which he described as ``freely exercised volition of the individuals concerned'' is simply draconian and cannot be sustained in modern democratic societies. However, ``unalloyed struggle for existence'' to prevent further accumulation of genetic load as an alternative to ``freely exercised volition'' seems even worse. An outright denial of both extremes without alternative solutions will lead to the unprecedented rise in incidence of complex diseases, as currently observed in developed countries.

Time-separated twinning has several attractive attributes that are absent in the techniques previously proposed to address this problem. First, it will facilitate prevention of further accumulation of genetic load because the genetic material used would be from the past. Second, surrogate parents are fully informed about the health status of an adult co-twin(s) of an embryo they consider for transfer (implantation). Thus, their educated choices will result in a significant reduction in many of the complex devastating childhood diseases such as asthma, autism, and type 1 diabetes, which have the highest concordance between younger MZ twins. This technique would lead to virtual eradication of diseases with simple Mendelian etiology, given that they are triggered by concrete causative variants and have very high concordance rates.

The Ethics Committee of the American Society of Reproductive Medicine (ASRM) has stated in its report that ``splitting one embryo into 2 or more embryos could serve the needs of infertile couples in several ways. As long as a couple is fully informed of the risk of such an outcome, there would appear to be no major ethical objection'' \cite{EmbryoSplittingEthics2004}. Such endorsement makes it easy to implement the procedure in IVF clinics already existing in many parts of the world. It is especially encouraging that half of the respondents in a recently conducted questionnaire were willing to accept belated twinning as an option for assisted parenthood \cite{discussingTwinningIssues2010}, because arguments against such procedures are frequently exaggerated and result from misconception of the existing reality \cite{discussingTwinningIssues2010,EthicsDuplication2000,GoodbyeDolly1997}.

It has been noted that naturally occurring MZ twins are socially accepted in the most extreme form of their manifestation where 2 (or even 3!) identically looking individuals of the same age live together with their parents \cite{twinsSweden1996,growingAsTwin1991}. For that reason, time-delayed twins should not confront many of the psychological barriers associated with the identical-age MZ twins \cite{spareOrIndividual1998}. The use of artificial twinning must be controlled by the government, as has been done for the use of donor eggs, donor spermatozoa, and surrogacy. For example, restrictions could be made that organs and tissues could only be regenerated from the donor's own totipotent cells.

The main advantage of the proposed method is that people affected by a hereditary disorder will still be able to raise healthy children considering that they borrow genetic material from the demi-embryo repository. Although modern fertility centers already offer surrogate parenthood with sperm and egg donors of superior health status, these procedures do not guarantee that a child conceived with the use of donated germinal material will also be healthy. Moreover, such options are very expensive considering the price of $\sim\$1,000$ per human egg together with the costs of all other associated procedures done in IVF clinics. Time-separated twinning has the potential not only to reduce substantially the costs of surrogate parenthood but also to allow more accurate health estimates of a future child than those based on the health of sperm and egg donors. The good health of multiple MZ co-twins, as compared to only one co-twin, would suggest even stronger concordance for the absence of a disorder for the particular demi-embryo.

As discussed in subsection {\em Type 2 diabetes}, the direct medical expenses associated with T2D in China alone were \$26.0 billion dollars in 2007 and are projected to increase significantly. According to our calculations presented in Table \ref{tab:prevention}, a 3.75-6.12-fold reduction in T2D could be achieved (excluding report \cite{DiabetesConcordanceFinland1992} with a suspiciously low concordance rate) following health status-guided belated twinning in a pure form. Preventive measures alone, as discussed in the China Da Qing Diabetes Prevention Study, can only reduce the incidence of T2D by 51\% and would require substantial lifestyle intervention \cite{diabetesInterventinChina2008}. Therefore, an approach that combines lifestyle intervention and belated twinning could significantly reduce disease incidence compared to both techniques used separately.

Concordance between MZ twins for T1D is smaller than for T2D. Therefore, health status-guided belated twinning could reduce the disease incidence no more than 2-fold (according to Table \ref{tab:prevention}), given that the steady rise of T1D affects mostly younger age patients (before age 10) among whom concordance is highest. Although this may appear to be a rather modest improvement, a 2-fold incidence reduction could provide great relief considering the devastating effect of the disease on the affected families.

If the concordances for ASD listed in Table \ref{tab:prevention} are indeed that high,
time-separated twinning could easily prevent 90\% of cases. Several lines of evidence suggest that these number are not exaggerated.
Based on separated MZ twins data heritability on schizophrenia could be as high as 0.80 \cite{HeritabilityIQ2004}.
Moreover, autism, bipolar disorder, and schizophrenia share a substantial number of genetic factors and show significant correlation \cite{probingGeneticOverlap2007}.

Unfortunately, not all diseases could be effectively prevented through time-separated twinning, including cancer. The data in Table \ref{tab:prevention} suggest that health status-guided belated twinning could reduce driver mutations and genetic predisposition for breast cancer by 1.14-fold, colorectal cancer by 1.17-fold, and prostate cancer by 1.25-fold. The causes of cancer are complex and include environmental conditions and some genetic predisposition factors in the form of inherited or {\em de novo} driver mutations and induced somatic variants.

\subsection*{Creation of demi-embryo repositories}

Currently, there are no extensive demi-embryo repositories in existence and time-separated twining in humans remains only an interesting theoretical possibility. Therefore, establishment of an ultra-long-term storage repository under government control would be a logical step forward. It might take several decades before any procedures involving belated twinning become ethically acceptable. During this time the repositories could collect a substantial number of demi-embryos from families willing to donate co-twins of their children. Under certain circumstances, such as the death of a child, parents should be able to request co-twin transfer from the repository if they decide such action would compensate their loss. Availability of autologous demi-embryos (the person's own totipotent cells) in the repository could revolutionize regenerative medicine and facilitate novel stem cell therapies. It appears that there are no simpler ways than artificial twinning to obtain autologous stem cells for therapeutic purposes. In a few decades from now the incidence of hereditary disorders in developed countries may simply become intolerable. As shown here, the incidence of allergies, ASD, and diabetes continues to rise in the younger generation. Future parents may consider embryo transfer from the repository as the only chance to have a healthy child.

The availability of demi-embryo repositories would create new possibilities for people living in regions affected by radioactive fallout.  The rates of birth defects in areas contaminated by radiation from the Chernobyl nuclear accident are higher than in uncontaminated regions of the Ukraine, suggesting that exposure to radiation {\em in utero} induces birth defects and represents a risk to human health \cite{radiationBirthDefects2010}. Therefore, assisted reproduction using demi-embryos from repositories in radiation-contaminated territories could eliminate the much-feared possibility of radiation-induced birth defects.

In considering the use of repositories there are 3 types of volunteer choices that could be made.

\begin{description}
  \item[Not affiliated:] People who wish not to be affiliated with the repository in any way.
  \item[People willing to deposit:] Couples who wish to have their own child and would like to have the co-twin in the cryoconservation repository. This choice offers several benefits such as having a genetically identical backup copy of their own child and having autologous totipotent cells that could be differentiated to other types of stem cells for future therapies. This procedure would involve hyperovulation, collection of oocytes with ultrasound-guided needle under local anesthesia, intracytoplasmic injection of partner sperm, several rounds of artificial twinning, and PGD to test for aneuploidy and Mendelian disorders. If PGD predicts no serious health reservations, the resulting demi-embryos could then be implanted and the remaining blastocysts vitrified. It has been mentioned that the main reason for the low pregnancy rate in women over 40 is the lack of implantation and the rate of loss of aneuploid embryos \cite{ChromosomalAbnormalitiesaAge1981,ChromAbnormality1983,OriginAneuploidy2012,Crow2000}. Therefore, collection of sperm and eggs at younger ages (assuming that couples are willing to use the repository later in their life) may improve the pregnancy rates of women in advanced age and reduce the incidence of chromosomal abnormalities associated with late parenthood \cite{Crow2000,NewBornMutations2012,Microsats2012}.

  \item[People willing to re-derive:] Couples diagnosed with hereditary health problems that could be inherited by their own child might consider transfer (implantation) from the repository of a demi-embryo with an adult healthy MZ sibling(s). This would significantly reduce the chances of a severe childhood disorder based on the previously established concordance between MZ twins. This procedure is more straightforward than embryo deposition and involves a simple embryo transfer.
\end{description}

The costs of long-term cryoconservation include the maintenance of the Dewar reservoir and timely replenishment of liquid nitrogen, which costs around \$0.06 per liter when purchased in bulk. The costs of cryoconservation per embryo will be negligibly low if a large number of blastocysts are simultaneously stored in a reservoir. Redundant storage should be arranged so that co-twins are deposited in at least 2 cryoconservation vessels at different secure locations.

H. J. Muller believed that it was possible to guide the evolution of mankind and
create a better allotment of positive qualities than would naturally occur \cite{InTheNameOfEugenics1985}.
He stated that selection artificially guided by concerned individuals expressing their free will
``is the only real solution, the only procedure consistent
with human happiness, dignity, and security'' and at the end
he concluded ``Strange as it seems, we can in that case both eat our cake and
have it'' \cite{Muller1950}.

\bigskip

\section*{Author's contributions}
AC conceived the study, performed the statistical analysis and wrote the manuscript.
LA participated in its design and coordination, helped to draft the
manuscript and have given final approval of the version to be published.
All authors read and approved the final manuscript.

\section*{Acknowledgements}
\ifthenelse{\boolean{publ}}{\small}{}
Special thanks to all the colleagues from the Beijing Institute of Genomics (BIG)
for helpful discussions that substantially improved the quality of this work.
Supported by grant 2011Y1SA09 from the Chinese Academy of Sciences Fellowship for
Young International Scientists and by grant 31150110466 from the National Natural
Science Foundation of China (NSFC).


\newpage
{\ifthenelse{\boolean{publ}}{\footnotesize}{\small}
 \bibliographystyle{bmc_article}  
  \bibliography{bmc_article} }     


\ifthenelse{\boolean{publ}}{\end{multicols}}{}

\end{bmcformat}
\end{document}